\documentclass[prb,twocolumn,showpacs,preprintnumbers,amsmath,amssymb]{revtex4}

\usepackage{graphicx}
\usepackage{dcolumn}
\usepackage{bm}

\begin{document}

\title[Short Title]{$^{55}$Mn NMR Study of the Field-Controlled Magnetic Phase Separation
in (La$_{0.25}$Pr$_{0.75}$)$_{0.7}$Ca$_{0.3}$MnO$_3$ with Different Oxygen Isotope
Content}

\author{A.Gerashenko}
\affiliation{Division of Physics, Graduate School of Science, Hokkaido University,
Sapporo 060-0810, Japan}\affiliation{Institute of Metal Physics, Ural Branch of
Russian Academy of Sciences, Ekaterinburg GSP-170, Russia}

\author{Y.Furukawa}
\affiliation{Division of Physics, Graduate School of Science, Hokkaido University,
Sapporo 060-0810, Japan}

\author{K.Kumagai}
\affiliation{Division of Physics, Graduate School of Science, Hokkaido University,
Sapporo 060-0810, Japan}

\author{S.Verkhovskii}
\affiliation{Institute of Metal Physics, Ural Branch of Russian Academy of Sciences,
Ekaterinburg GSP-170, Russia}

\author{K.Mikhalev}
\affiliation{Institute of Metal Physics, Ural Branch of Russian Academy of Sciences,
Ekaterinburg GSP-170, Russia}

\author{A.Yakubovskii}
\affiliation{Russian Research Centre "Kurchatov Institute", Moscow
123182, Russia}

\smallskip
\begin{abstract}
An influence of the $^{16}$O-$^{18}$O isotope substitutions on magnetic state of
perovskite-type manganite (La$_{0.25}$Pr$_{0.75}$)$_{0.7}$Ca$_{0.3}$MnO$_3$ is studied
by $^{55}$Mn NMR. Successive cycling with an isochronal exposure at different magnetic
fields up to $H=8T$ is used to study the field-induced transition from
antiferromagnetic insulating (AFI) state to the ferromagnetic metal (FMM) one in the
$^{18}$O-enriched sample. After exposure at $H>H_{cr}\sim~5.3T$ the NMR spectrum of
the $^{18}$O-sample evidences for magnetic phase separation (PS) resulted in the
coexisting AFI and FMM domains. Further increase of exposing field leads to a
progressive growth of the FMM phase at the expense of AFI domains. Its relative
fraction can be controlled by external magnetic field and the resulting magnetic
structure in the PS region is discussed. Anomalous $T-$dependence of the $^{55}$Mn
nuclear spin-lattice relaxation rate is revealed in the FMM state of both $^{16}$O-
and $^{18}$O-enriched samples. The manifestation of the Pr magnetic ordering at $T
\sim40K$ is considered.
\end{abstract}

\maketitle
\smallskip
\section{Introduction}

Perovskite-type manganites R$_{1-x}$M$_{x}$MnO$_3$ (R = La, Pr is trivalent rare-earth
ion) are subjected to extensive studies after observation of a "colossal" negative
magnetoresistance (CMR) effect for $0.2<x<0.4$. The CMR effect relates in physics
close to phase transition from the charge ordered antiferromagnetic insulating (CO
AFI) state to the ferromagnetic metal (FMM) one \cite{Coey}. Thermal hysteresis
revealed in transport and magnetic properties of these compounds evidences the first
order transition accompanied by the phase separation (PS) into CO AFI and FMM domains
\cite{Nagaev}. An ionic state of Mn determines unambiguously its spin configuration,
substantial polaronic effects \cite{Khomskii} and Jahn-Teller type lattice distortions
\cite{Zhou} in sublattice of the MnO$_6$ octahedra and defines in many respects the
microstructure of magnetic state in a low temperature phase.

The (La$_y$Pr$_{1-y}$)$_{0.7}$Ca$_{0.3}$MnO$_3$ manganite is one of the most
convenient systems for studying the PS phenomenon. Recently \cite{Babushkina_N391,
Babushkina_PRB60} it was shown that the ground state of electron system becomes
extremely unstable at $y\sim0.25$ and even the isotopic substitution of $^{16}$O by
$^{18}$O influences significantly the transport and magnetic properties of this
compound. On cooling down in zero external magnetic field (ZFC), the
(La$_{0.25}$Pr$_{0.75}$)$_{0.7}$Ca$_{0.3}$MnO$_3$ sample with $^{16}$O isotope
(referred below as LPCMO$^{16}$ ) shows the successive transitions to the CO state at
$T<T_{co}=180K$, to the AF state below $T_N=150K$ and finally to the FMM one below
$T_c=120K$. On the other hand, the $^{18}$O-enriched sample (LPCMO$^{18}$) remains in
the AFI state down to the very low temperature under ZFC conditions. However, this AFI
ground state is extremely unstable and can be easily transformed to the FMM one with
the same saturation magnetization as in the LPCMO$^{16}$ by applying external magnetic
field above the critical value $H_{cr}$.

The detailed magnetic phase diagram of LPCMO$^{18}$ in a wide range of $H$ and $T$ has
been reported as result of magnetization studies \cite{A.Y&K.K}. The AFI-FMM phase
composition in the PS region above $H_{cr}(T)$ is shown to be controlled by magnetic
field. Any prescribed ratio of AFI to FMM phases can be obtained in the PS region and
might be frozen by decrease of the magnetic field below the critical value $H_{cr}$.
The new phase composition depends on neither time nor magnetic field variations below
$H_{cr}$. The microscopic evidence of the phase transition from AFI to FMM state in
LPCMO$^{18}$ was obtained from $^{139}$La NMR studies at $5K$  \cite{A.Y&K.K}.  Two
well-resolved $^{139}$La NMR lines corresponding to AFI and FMM domains were clearly
observed and their relative intensities determine directly the fraction of the both
phases in the PS region.

However, $^{139}$La NMR spectra are not informative in concern to the short-range CO
of the Mn ions in the AFI phase. Indeed the local magnetic field $^{139}H_{loc}$
probed by the $^{139}$La is mainly due to the overlap of La(6s) and Mn($t_{2g}$)
orbitals and the contribution of holes located at the $e_g$-orbital of Mn$^{3+}$ ions
is greatly reduced in the insulating state \cite{Yoshinari}. Thus the position of the
$^{139}$La NMR signal in the AFI state is not sensitive to the difference of the
Mn$^{4+}$(d$^3$)/Mn$^{3+}$(d$^4$) valence state of the nearest Mn ions. In sharp
contrast, one may expect that $^{55}$Mn NMR study allows to obtain more detailed
information about the charge and magnetic states of Mn ions in the AFI phase.

In this paper, we present the results of the $^{55}$Mn NMR studies of the
field-controlled PS in (La$_{0.25}$Pr$_{0.75}$)$_{0.7}$Ca$_{0.3}$MnO$_3$ with
different oxygen isotope content to get better insight in the microstructure of the
AFI ground state. The temperature dependence of nuclear spin-lattice relaxation rate
is studied in the FMM state with special care to find the microscopic evidence for
magnetic ordering of Pr ions.

\section{EXPERIMENTAL}
The ceramic sample preparation and isotope enrichment procedures
were described in detail elsewhere \cite{Babushkina_N391}. The NMR
measurements were performed with a home-built pulse phase-coherent
NMR spectrometer operated in frequency range up to $450MHz$ using
spin-echo technique. The NMR spectra were obtained by measuring at
each frequency an intensity of the Hahn spin-echo signal. The
width of a $\pi/2$ rf-pulses does not exceed $(1.5- 2)\mu s$. The
amplitude of the exciting rf-pulses was optimized for the maximum
of echo signal in measuring each individual line in the NMR
spectrum. Any variations of the receiver gain including
rf-coupling with resonant circuit were taken into account in
measurement of the line intensity using an additional calibration
rf-pulse with fixed amplitude formed in the rf-coil after
echo-signal at each frequency point. The original probe-head
designed for $^{55}$Mn NMR allowed to measure spectra in the range
of $180-450MHz$ using a single rf-coil. The $^{55}$Mn NMR spectrum
measurements were performed in zero magnetic field (ZFNMR) and in
external magnetic field up to $8T$  at $T=1.5K$. The field-cycling
(fc) procedure $(ZF\rightarrow~H_{fc}\rightarrow~ZF)$ was
performed as follows: after ZFC the external magnetic field was
increased up to $H_{fc}$ and was kept fixed for about
$t_{exp}\sim~20~min$ to exclude the transition effects. Then the
magnetic field was switched off and ZFNMR spectrum was measured.

\section{RESULTS AND DISCUSSIONS}

\begin{figure}[bp]
\includegraphics[width=0.3\textwidth,viewport=85 0 500 800]{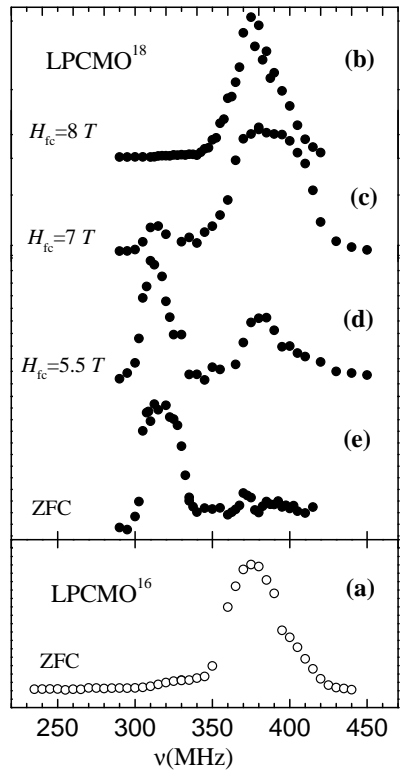}
\caption{ $^{55}$Mn spin echo spectra of LPCMO$^{18}$ (solid
circles) and LPCMO$^{16}$ (open circles). Spectra were measured at
the polycrystalline powder after primary ZFC cooling down to
$T=1.5K$ and subsequent isochronal magnetic field circling
$H=0>H_{fc}>H=0$ with $H_{fc}=0T \longrightarrow$(a) and (e),
$8T\longrightarrow$(b), $7T\longrightarrow$(c),
$5.5T\longrightarrow$(d). The line is a guide for eye.}
\label{fig1}
\end{figure}

Figure 1(a) shows the $^{55}$Mn ZFNMR spectra in the FMM state in LPCMO$^{16}$ after
ZFC. It presents a single line peaked near $\nu\approx~380MHz$. The line shows a
strong rf-enhancement $\eta\approx~100$ which is typical for ordered FM. On the other
hand, in the AFI state of the LPCMO$^{18}$ sample after ZFC, the observed spectrum is
quite different, as shown in Fig.1(e). The main peak is observed around $317MHz$ where
a large rf-power is needed for the signal detection in contrast to the signals from
the FMM state, while the small peak around $380MHz$ with a large rf-enhancement is
also detected. Besides the rf-enhancement factors the other characteristics of NMR
signals for these two peaks are also completely different. Nuclear spin-spin
relaxation time $T_2=10(5)\mu s$ $(T=1.5 K)$ and nuclear spin-lattice relaxation time
$T_1=2.9ms$ at the main peak are much shorter than the corresponding $T_2  \sim 100\mu
s$ and $T_1 \sim 1s$ observed at the small one. These short relaxation times are
typical for AFI state in manganites \cite{Allodi_PRL81}. Thus we attribute the peak
around $380MHz$  to $^{55}$ Mn NMR in the FMM phase while that around $317MHz$ is
assigned to the AFI phase.

Figures 1(b,c,d) show the $H_{fc}$-dependence of the $^{55}$Mn spectrum measured with
the field cycling method described above. As shown in the figures, with increasing
$H_{fc}$, the peak around $317MHz$ originated from the AFI phase gradually disappears,
while a relative intensity of the peak around $380MHz$ (FMM phase) grows up. Finally,
the spectrum in the LPCMO$^{18}$ after the exposing to $H_{fc}=8T$ (Fig. 1(b)) becomes
very similar to that in the FMM state of the LPCMO$^{16}$, Fig.1(a). From the
comparison of FMM signal intensities between the virgin (ZFC) and the final (exposing
to $H_{fc}= 8T$) states the minor FMM phase in the virgin LPCMO$^{16}$ sample can be
estimated at most a few percent. It's worth mentioning that only intensities not the
resonance frequencies of the peaks changed during this field cycling procedure, which
proves that only concentrations of the two defined magnetic phases were changed.

\begin{figure}[bp]
\includegraphics[width=0.5\textwidth,viewport=40 250 850 800]{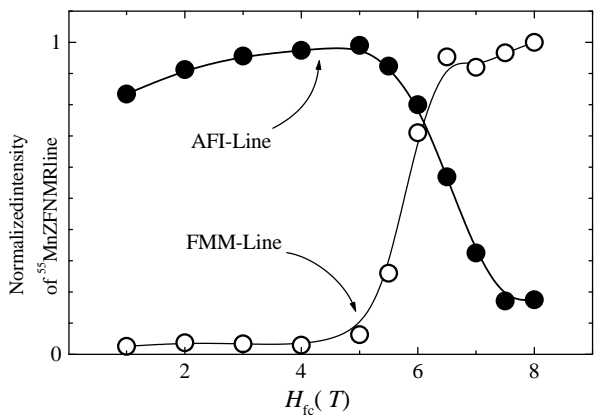}
\caption{The relative intensity of $^{55}$Mn ZFNMR line in the AFI domains ($\nu \sim
317 MHz$, solid circles) and the FMM domains ($\nu \sim 380MHz$, open circles)
measured at T = 1.5 K for LPCMO$^{18}$ after subsequent isochronal magnetic field
cycling at $H_{fc}$. The intensities are normalized to the corresponding maximums.}
\label{fig2}
\end{figure}

The amount of AFI or FMM phases is proportional to intensity of the corresponding
ZFNMR line measured at $T=1.5K$ after subsequent field cycling at different $H_{fc}$.
The relative intensities of the AFI($\bullet$) or FMM($\circ$) lines normalized to
their maximum are shown in Fig.2 for the virgin (ZFC) LPCMO$^{18}$. With increase of
$H_{fc}$ the amount of AFI domains increases slightly below $H_{fc}\sim~5T$,  then
starts to decrease nearly to zero at $H_{fc}=8T$. On the other hand, the amount of FMM
domains increases strongly above $H_{fc}\sim~5T$ and saturates at $H_{fc}\sim~7T$. In
the region between $\sim~5.5T$ and $7T$, the AFI and FMM domains coexist in
LPCMO$^{18}$, indicating the microscopically inhomogeneous magnetic phase separation.
These $^{55}$Mn-NMR results provide a microscopic confirmation of the inhomogeneous PS
in LPCMO$^{18}$ in the region where a detailed balance of the volume fractions of the
AFI and the FMM phases can be controlled by external magnetic field. The present
results are consistent with $H-T$ magnetic phase diagram obtained by magnetization
measurements and $^{139}$La NMR studies \cite{A.Y&K.K}.

\begin{figure}[bp]
\includegraphics[width=0.35\textwidth,viewport=10 10 515 815]{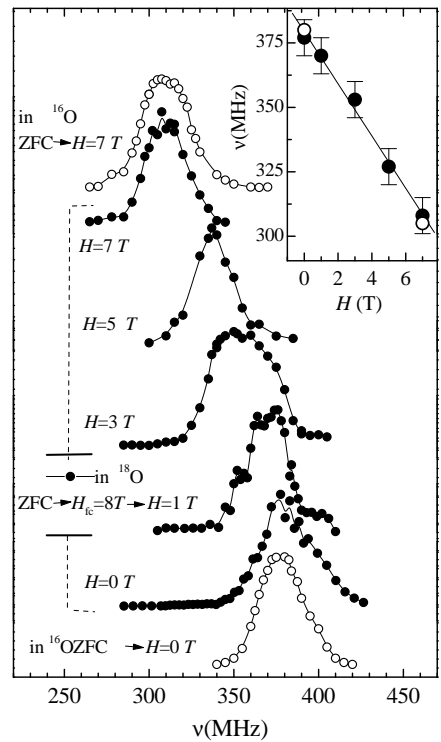}
\caption{ $H$-dependence of $^{55}$Mn spin echo spectra measured at $T=4K$ for FMM
phase of LPCMO$^{16}$   (open circles) and LPCMO$^{18}$ (solid circles) samples. The
ZFC LPCMO$^{18}$ sample was transformed to FMM state by exposing at $H_{fc}=8T$. The
inset shows the frequency of the NMR line peak $\nu$ vs $H$, a slope of solid fitting
line corresponds to $\Delta\nu/\Delta~H=10.1MHz/T$.} \label{fig3}
\end{figure}

In order to shed light on more details of the electronic and
magnetic states of Mn ions in FMM phase, we have investigated the
external magnetic field dependence of the $^{55}$Mn NMR spectra.
Fig.3 shows the $H$-dependences of $^{55}$Mn NMR spectrum measured
in LPCMO$^{16}$ at $T=4K$ (open circles) and of the spectrum for
saturated FMM LPCMO$^{18}$ corresponding to Fig.1(b). With
increase of $H$ the peak frequencies for both samples shift to
lower frequency according to the relation $\Delta
\nu(MHz)\sim10.1H(T)$ as shown in the inset of Fig.3. This value
is in agreement with the gyromagnetic ratio of $^{55}$Mn nucleus
$(^{55}\gamma/2\pi=10.50T/MHz)$ within our experimental accuracy.
As the hyperfine field $H_{hf}$ at the Mn sites is mainly
originated from the core-polarization contribution, its direction
is opposite to that of the Mn spin moment. This result suggests
that all the Mn spin moments in the FMM state aligned along the
$H$ direction without any canting. $H_{hf}$ is proportional to
$A_{hf}\langle S \rangle$, where $A_{hf}$ is hyperfine coupling
constant and $\langle S \rangle$ is spin moment. Combining the
experimental value of $H_{hf}=36T$, $A_{hf}\approx -100kOe/\mu_B$
\cite{A.Freeman} and the magnetization data measured in the same
sample LPCMO$^{16}$, the average magnetic moment on the Mn site is
estimated as $3.6\mu_B$. This value suggests that the valence of
Mn ions in the FMM state can be given to Mn$^{3.6+}$ in high spin
states in ionic approximation; namely some averaged value between
Mn$^{4+}$ (S=3/2) and Mn$^{3+}$ (S=2). This in turn suggests that
the electronic state of 0.6 electrons on the $e_g$-orbitals in FMM
state must be responsible for metallic conductivity. It should be
pointed out that the observation of only one component of the Mn
NMR spectrum in the FMM state indicates that the inverse of the
life-time of the electron spins is higher than the NMR frequency
(in the exchange narrowing limit between Mn$^{4+}$ (S=3/2) and
Mn$^{3+}$ (S=2)).

On the other hand, for the CO AFI phase, we would expect two
$^{55}$Mn NMR signals from Mn$^{4+}$ (S=3/2) and Mn$^{3+}$ (S=2)
ions, because $e_g$ electrons of Mn$^{3+}$ are expected to
localize in the AFI state. However, we observed only one signal
around $317MHz$ corresponding to $H_{hf}\sim~30.2T$. The magnetic
moment of the Mn ions in the AFI state is suggested to be smaller
than that of the Mn$^{3.6+}$ ions in the FMM state. Furthermore,
the value of $H_{hf}\sim~30.2T$ is well consistent with a
theoretical estimation \cite{T.Kubo} of the on-site hyperfine
magnetic field for the Mn$^{4+}$ ion in crystal field of
octahedron symmetry $H_{hf}$ (Mn$^{4+})=30.5T$. Hence, thus this
signal can be assigned to Mn$^{4+}$ ions in the CO AFI phase.

As for the NMR signals from Mn$^{3+}$ ions in the CO AFI state, we have not succeeded
to find them in the frequency range of $320-450 MHz$ even with the shortest available
delay time ($5\mu s$) between rf-pulses. This might be due to a much shorter $T_2<1\mu
 s$ for Mn$^{3+}$ ions compared to $T_2 \sim 10\mu s$ for Mn$^{4+}$ ions in the CO AFI
state. A similar negative result was obtained in the $^{55}$Mn NMR spin-echo studies
of the CO AFI state of La$_{0.5}$Ca$_{0.5}$MnO$_{3}$ \cite{Allodi_PRL81} and
Pr$_{0.5}$Sr$_{0.5}$MnO$_3$ \cite{Allodi_PRB61}.

Finally we consider peculiarities of the $^{55}$Mn nuclear spin-lattice relaxation
rate $(T_1^{-1})$ probing the fluctuations of local magnetic fields in the FMM ordered
state of LPCMO$^{16}$ (ZFC) and LPCMO$^{18}$ ( FC at $8T$) samples. The
$T_1$-measurements were performed in magnetic fields $H=1;2T\ll~H_{cr}$ using the
saturation-recovery method. The recovery of nuclear magnetization $M_z(t)$ of
$^{55}$Mn (I=5/2) to its thermal equilibrium value $M_z(\infty)$ after the applied
saturating comb of pulses is given by \cite{Narath_PR162}
\begin{eqnarray}
\frac{(Mz(\infty)-Mz(t)}{Mz(\infty)} =\frac{1}{35}exp(-\frac{t}{T_1}) + \nonumber \\
\frac{8}{45}exp(-\frac{6t}{T_1}) + \frac{50}{63}exp(-\frac{15t}{T_1})
\end{eqnarray}
The expression (1) was used for fitting the obtained $M_z(t)$ data.

\begin{figure}[bp]
\includegraphics[width=0.35\textwidth,viewport=35 80 520 680]{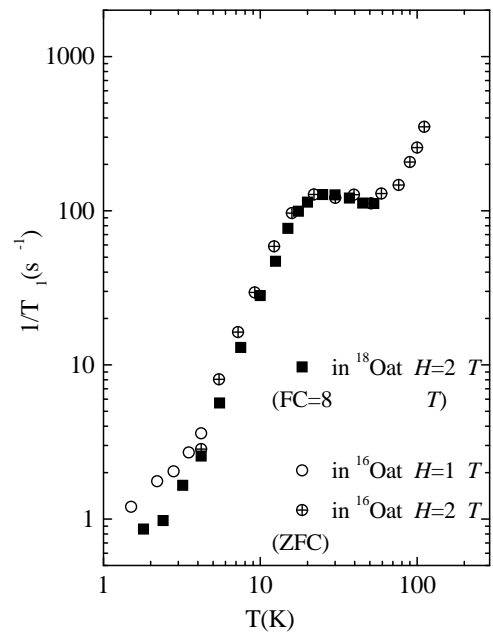}
\caption{ Temperature dependence of $^{55}$Mn nuclear spin-lattice relaxation rate
$T_1^{-1}$ measured in the FMM state of LPCMO$^{16}$ and LPCMO$^{18}$ samples. The ZFC
LPCMO$^{16}$ sample was measured in $H=1T(\circ)$ and $2T(\oplus)$, the LPCMO$^{18}$
one was FC in $8T$ and $T_1$-measurements were performed in $H=2T(\blacksquare)$.}
\label{fig4}
\end{figure}

The $T$-dependence of $1/T_1$ for both ZFC LPCMO$^{16}$ (circles) and LPCMO$^{18}$
(solid squares) samples cooled down at $H=8T$ is presented in Fig.4.  It's seen, that
in the FMM state the temperature behavior of $1/T_1$ is similar for both samples. With
increasing $T$, $T_1^{-1}$ increases strongly up to $T=20K$, have a plateau in the
range of $T=20-50K$, and increases again on approaching the transition from the FMM to
the AFI state at $T_c=120K$. This finding clearly indicates that low frequency
fluctuating parts of the local magnetic fields at the Mn sites in the FMM state are
not noticeably influenced by oxygen isotope substitution.

As expected at elevated temperatures near $T_c$ the critical enhancement of magnetic
fluctuations at the Larmor frequency of $^{55}$Mn nuclei is responsible for a rather
strong growth of $(T_1^{-1})\sim T^3$. On the other hand, $T_1^{-1}$ behavior at low
temperature is quite different compared to the exponential rise $^{55}T_1^{-1} \sim
exp(aT)$ \cite{Allodi_PRL81,Savosta_PRB59} observed for several FM manganites with
concentration of mobile holes covering the CMR region of coexisting FM order and
metallic conductivity. Similar hump in the temperature dependence of $^{139}T_1^{-1}$
was also revealed in measurements of the $^{139}$La nuclear spin-lattice rate
performed below $100K$ for the FM line of $^{139}$La of the LPCMO$^{18}$ sample FC at
$8T$ \cite{Furukawa}. It is instructive to note that in ferromagnetic metals the
dominating spin-wave contribution to $T_1^-1$ via the two-magnon scattering process
should lead to the power-law dependence $T_1^{-1}(T)\sim T^\beta$ with $\beta$ ranging
from 2.5 to 1.5 with an increase of the magnon damping \cite{Irkhin_PRB60}.

These observations imply the unique source of fluctuating magnetic fields which
contributes to nuclear spin-lattice rate of $^{139}$La and $^{55}$Mn in FMM state of
(La$_{0.25}$Pr$_{0.75}$)$_{0.7}$Ca$_{0.3}$MnO$_3$. We incline to conclude that the
anomalous $T_1$-behavior observed at low temperature is originated from the
fluctuating local magnetic field transferred from Pr whose spin fluctuations should be
increased greatly near the temperature of the Pr magnetic ordering. Indeed the
temperature range of $20 \sim 50K$ for the observed plateau of $1/T_1$ includes itself
the Curie temperature for Pr magnetic ordering $T_{c,Pr}(H=0)\sim40K$ as followed from
the magnetic $H-T$ phase diagram \cite{A.Y&K.K} and neutron diffraction studies
\cite{Balagurov_PRB60} reported for (La$_{0.25}$Pr$_{0.75}$)$_{0.7}$Ca$_{0.3}$MnO$_3$
with the very same chemical composition. To our knowledge, the present $^{55}T_1^{-1}$
results represents the first NMR evidence of the Pr magnetic ordering. An absence of
divergent-like behavior of $T_1^{-1}$ and rather large width of the plateau indicate
that magnetic order of Pr occurs as in spin-glass materials. It's quite natural since
the atomic disorder in the Pr/La sublattice should provide a rather large distribution
of the exchange parameters between the localized electron spins of Pr.

In conclusion, the $^{55}$Mn NMR spectra of
(La$_{0.25}$Pr$_{0.75}$)$_{0.7}$Ca$_{0.3}$MnO$_3$ show at a microscopic level that the
AF order among Mn ions in the LPCMO$^{18}$ sample at low temperatures is a metastable
magnetic state. After cycling of external magnetic field above $H_cr \sim 5T$ the
field- induced phase transition develops through the nucleation of the FMM phase at
the expense of AFI domains. An upper boundary of the AFI-FMM phase separation region
is established in the field- cycle NMR experiments and equals to $7.5T$ at $T=1.5 K$.
The $^{55}$Mn NMR spectrum of the LPCMO$^{18}$ sample cooled in the higher magnetic
field shows that its magnetic state is just the same as for LPCMO$^{16}$, i.e. a
long-range FMM. Finally, it's proposed that transferred magnetic coupling between the
Mn and Pr spins should be involved into consideration of the unusual low temperature
phase diagram of this manganite.

\begin{acknowledgments}
This study was supported in part by a Grant-in-Aid for scientific research of the
Ministry of Education, Culture, Sports, Science and Technology of Japan (A.G.), by
Russian Fund for Basic Research under Projects No 99-02-16975, 02-02-16357 and by CRDF
Grant RP2-2355-MO-02.
\end{acknowledgments}

\end{document}